\documentclass{aastex}          
\usepackage{spr-astr-addons}    
\usepackage{url}

\RequirePackage{color}
\begin{document}
%
\title{Temporal offsets among Solar activity indicators}

\shorttitle{<Short article title>}
\shortauthors{<Ramesh, K.B. and Vasantharaju, N.>}

\author{Ramesh, K.B.\altaffilmark{}} 
\affil{Indian Institute of Astrophysics,\\
Bangalore,India 560034\\
kbramesh@iiap.res.in}
\and 
\author{Vasantharaju, N.\altaffilmark{}}
\affil{Indian Institute of Astrophysics,\\
Bangalore,India\\
vasanth@iiap.res.in}



\begin{abstract}
Temporal offsets between the time series of solar activity indicators provide important clues regarding the physical processes responsible for the cyclic variability in the solar atmosphere. Hysteresis patterns generated between any two indicators were popularly used to study their morphological features and further to understand their inter relationships. We use time series of different solar indicators to understand the possible cause-and-effect criteria between their respective source regions. Sensitivity of the upper atmosphere to the activity underneath might play an important role in introducing different evolutionary patterns in the profiles of solar indicators and in turn cause temporal offsets between them.   Limitations in the observations may also cause relative shifts in the time series.
\end{abstract}


\section{Introduction}   
Almost every parameter derived from solar observations on longer time scales depict unambiguously 11-year solar cycle. The shape of the solar cycle, however, differs not only from the  patterns represented by different parameters, but also from cycle to cycle as represented by an independent parameter. This is because the evolution of different solar indices are not always similar (Kane, 2002). In particular, at solar maximum, often dual peaks are seen and their relative heights are not similar. Comparing two such indices of dissimilar patterns cause hysteresis (loop like structure) like phenomena. Hysteresis phenomena in solar activity is extensively studied  \citep{don91,1993ApJ...404..805B,bach94,2002cosp...34E.513O,bro95,jim98,dor99,2003JGRA..108.1379K,2011SoPh..269..451K,2012SoPh..276..407S,2012SoPh..281..839O} over the past two decades.  The hysteresis phenomenon in solar activity was examined in the perspective of storage and subsequent release of energy in the solar atmosphere \citep{2001ApJ...557..332W} while it was interpreted in terms of delayed process of active regions evolving from the photosphere upwards \citep{bach94}. \citet{bach04} do not  consider this as hysteresis phenomena because hysteresis implies a cause-and-effect relationship that depends on the history of the cause, whereas, all the solar activity indices are ultimately the effects of a common magnetic dynamo mechanism. Therefore, a comparison between two solar indices would provide a relative delay, if any, between the evolutionary patterns of those parameters and not with their respective driving force.  It is widely accepted that this driving force exists  in the solar interior and is produced by the solar dynamo. Sunspots, representing largely the strong field regions, are the prominently known feature that can throw light on the dynamo processes (Charbonneau, 2010).  We attempt to use this criteria to analyse the temporal offsets between the full-disk solar indices in order to understand their solar cycle variability.

\section{Data}  
Studies on long term variations of the solar activity require continuous data lasting several decades. Data base developed at ftp://ftp.ngdc.noaa.gov/STP/  is a great source of such long-term data  collected through variety of observations at different locations all over the globe. We use international sunspot numbers (SSN), sunspot area (SA), solar radio flux (F10), coronal green line index (CI), solar flare index (FI), background x-ray flux (XBF), total solar irradiance (IRR), calcium plage index (CaI), magnesium core to wing ratio (MgII), and occurrence rate of coronal mass ejections (nCME) in the present analysis.  

For solar cycle 21 that progressed during 1976 and 1986 we use SSN, SA,F10, FI, XBF, CaI and CI time series.  We use SSN, SA, F10, FI, XBF, CaI, CI, IRR, and MgII for the solar cycle 22 (1986 through 1996) and  SSN, SA, F10, FI, XBF, CI, IRR, MgII, and CME for the solar cycle 23  (1996 through 2007).

\subsection{Sources of data}  

SSN -	ftp://ftp.ngdc.noaa.gov/STP/ space-weather/ solar- data/solar-indices/sunspot-numbers/ international/listings/\\
SA    - 	http://solarscience.msfc.nasa.gov/greenwch.shtml\\
IRR  - 	ftp://ftp.pmodwrc.ch/pub/data/ irradiance/ composite\\ 
nCME - http://cdaw.gsfc.nasa.gov/cme\_list/\\
XBF – 	http://goes.ngdc.noaa.gov/data/avg\\
CaI -  	ftp://ftp.ngdc.noaa.gov/stp/ solar\_data/ solar \_calcium / index/McMath/docs/\\
FI    -	ftp://ftp.ngdc.noaa.gov/stp/ solar\_data/ solar flares/index \\
F10- ftp://ftp.ngdc.noaa.gov/stp/space-weather/ solar-data/solar-features/solar-radio/noontime-flux/ penticton/penticton\_adjusted/listings/\\
CI - 	ftp://ftp.ngdc.noaa.gov/stp/ solar\_data/ solar \_corona/index/\\
MgII - ftp://ftp.ngdc.noaa.gov/stp/solar\_data/solar \_UV\\

\section{Analysis}  
During the 21st solar cycle the background X-ray flux (XBF) data available for our analysis is monthly averages of daily values \citep{1988AdSpR...8...67W}. Daily background X-ray flux  is the minimum of three eight-hourly averages obtained from 5 min cadence GOES X-ray flux values. Monthly averaged XBF is computed using daily background X-ray flux. We compute similar time series of all the parameters considered in this work using their respective daily values. Further in order to produce smoother patterns of the time profiles, we apply 13-month box-car smoothing to monthly averaged values of each indicator. Before applying the 13-month smoothing, we have applied 5-month smoothing also to avoid ambiguity in identifying the occurrence of maximum among tiny multiple peaks around that time \citep{2012SoPh..276..395R}.The temporal variation of all the considered solar indicators for the three solar cycles is shown in Fig. 1.
\begin{figure}[hbtp]
  \centering
  \includegraphics[width=\columnwidth]{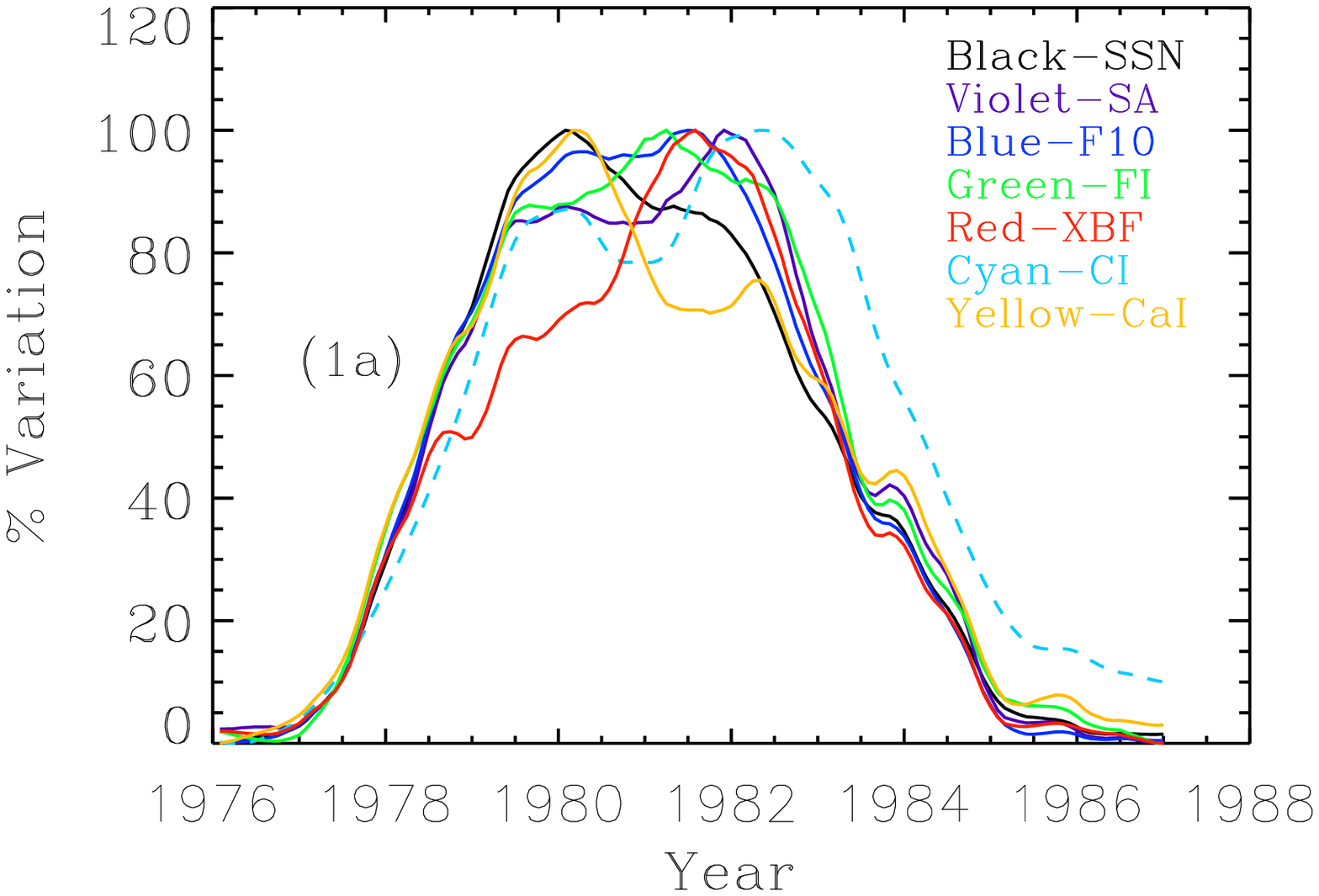}  
 \includegraphics[width=\columnwidth]{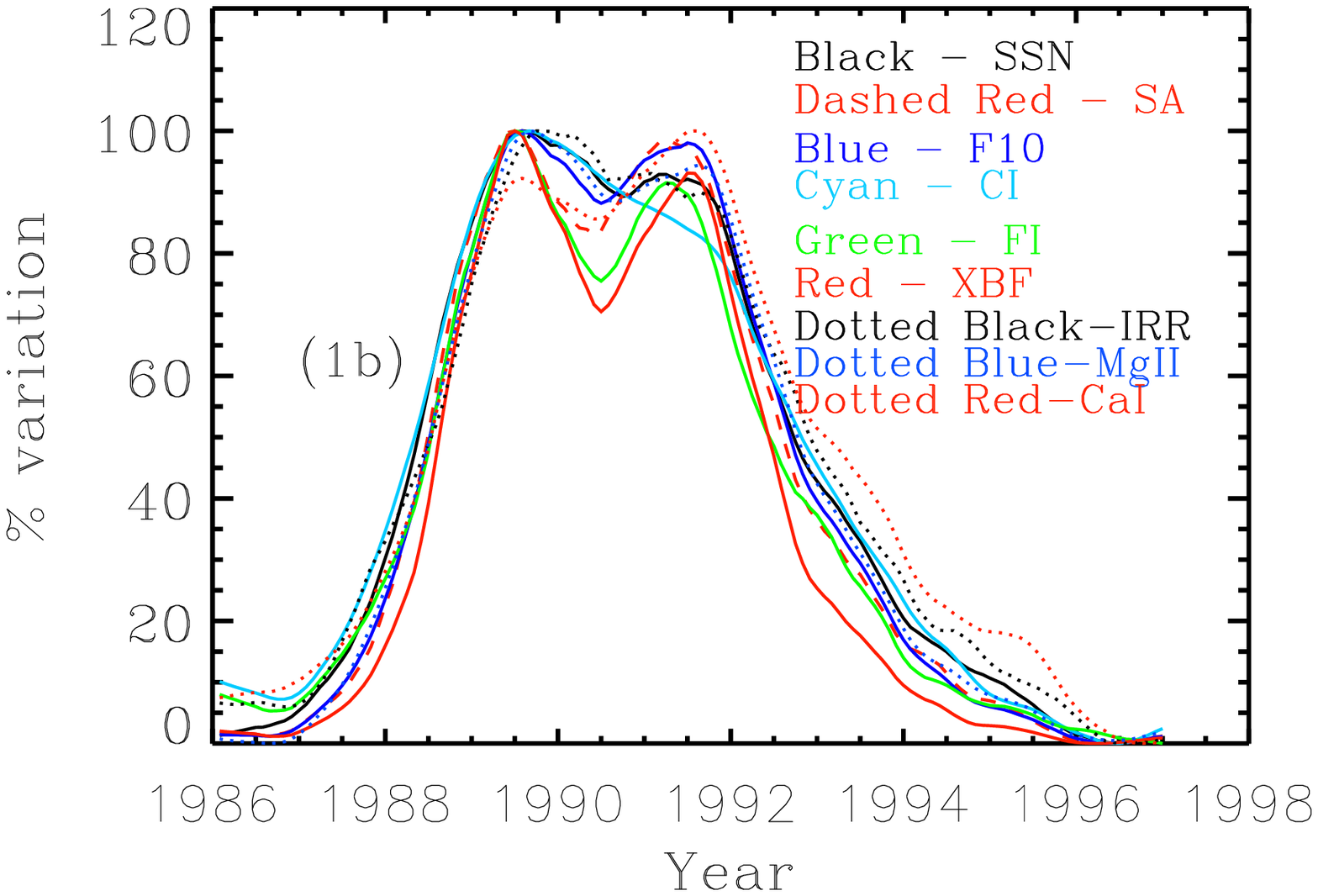} 
 \includegraphics[width=\columnwidth]{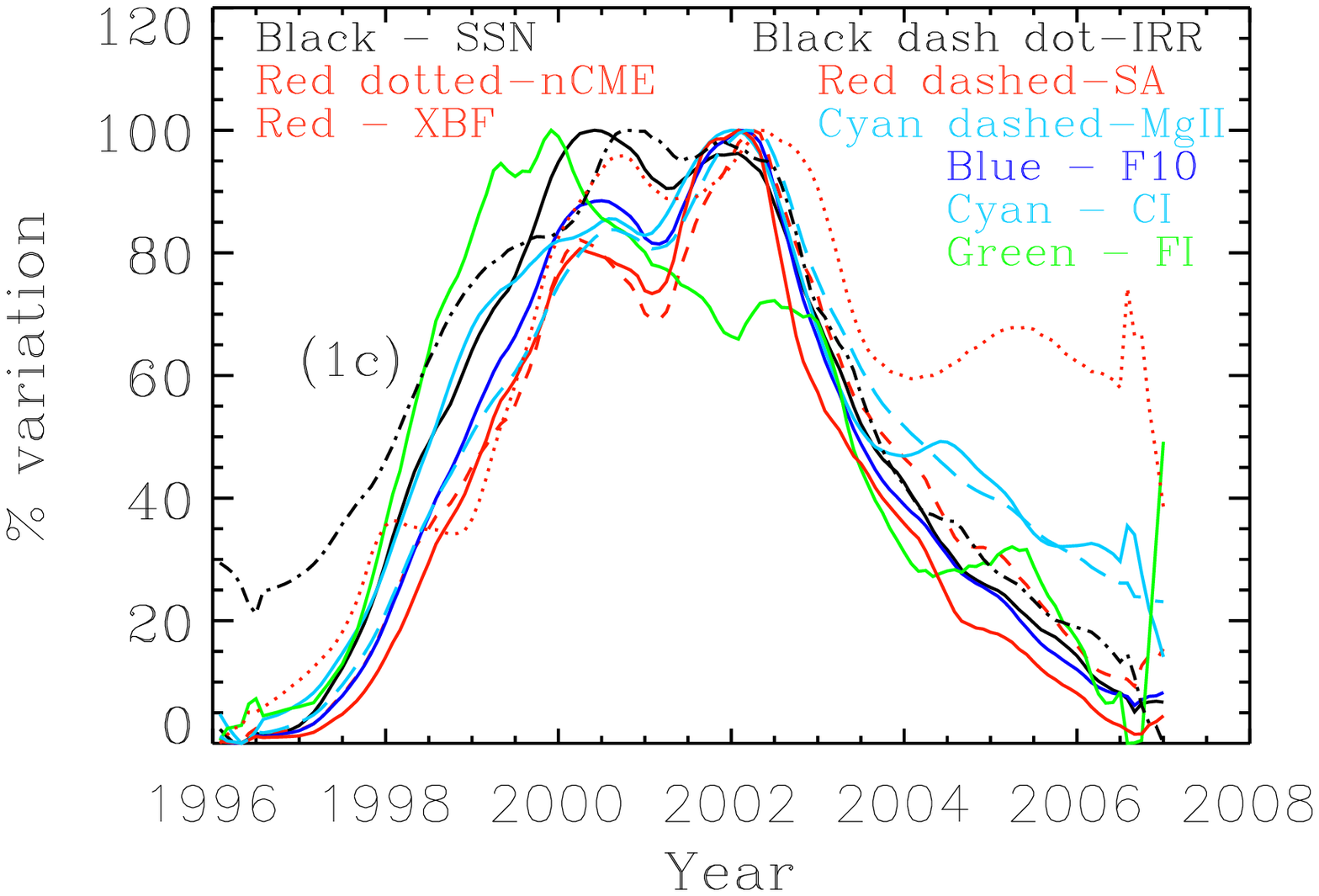}
  \caption[]{\label{fig:solar indices} %
Time evolution of smoothed solar indices for solar cycles 21 (Fig. 1a), 22 (Fig. 1b) and 23 (Fig. 1c)
 }
\end{figure}
 The ordinate represent the percentage variation\begin{equation}
 $$ \frac {Y_{i} -Y_{min}} {Y_{max} - Y_{min}} \times 100 $$
 \end{equation} of individual parameter. This procedure is adopted to  compare the variability in the long-term strong field active component, also known as slowly varying component (the S component) of the solar cycle. For example the 10.7 cm flux (F10) varies with solar activity while the quiet Sun component remains around 66 SFU (Solar Flux Units – $10^{-22}$ W $m^{-2}$ $Hz^{-1}$).The solar cycle related s-component seems to  vary from 66 to more than 200 SFU \citep{1990SoPh..127..321T}. The minimum value of each indicator remains constant within 10\% from one solar minimum to the other and defines the quiet-Sun background of the indicator \citep{1984ApJ...282..776S}.   Therefore we offset all the time series to a common value (Zero) by subtracting the respective minimum (quiet-Sun background of the indicator)  and then normalize to  the respective maximum in order to bring all the time series to a common scale and represent in terms of percentage variation as explained above.  This helps scrutinizing and comparing the evolution of time series for their temporal offsets, and temporal relationships, if any, in the overall rise and fall of the indices.

Figure 1 depicts unambiguously the popular 11-year period in all the three solar cycles.  During solar cycle 21 (Fig. 1a) the first part of the ascending and second part of the descending phases seem to be similar in all the considered parameters. The evolutionary pattern of XBF seems to differ from the other indices to a large extent in the second part of the ascending phase. The evolutionary pattern in the first part of the descending phase seems to be similar in all the parameters although they do not coincide. Pattern in the maximum phase of the cycle is more complicated. Except in CI and CaI, the double humped pattern is indiscernible in SSN, SA, F10, FI, XBF.  

During the cycle 22 (Fig. 1b)  the time profiles of all the parameters show similar trends in both ascending as well as in descending phases. The variability during maximum phase of the cycle  do not coincide with each other even though the double humped pattern is clearly seen in all the parameters considered.

Cyclic variability during solar cycle 23 (Fig. 1c) is more complicated, particularly the pattern representing the CME occurrence rate,  FI and IRR.  Ascending and descending phases of FI looks similar to those of SSN and SA. However, the occurrence of maximum of FI seems to deviate to a large extent with respect to the sunspot numbers and area. Occurrence rate of CMEs show delayed  pattern when compared to that of SSN and SA.  IRR deviates to a greater extent compared to  the other considered parameters in the ascending phase. 
\begin{figure}[hbtp]
  \centering
  \includegraphics[width=\columnwidth]{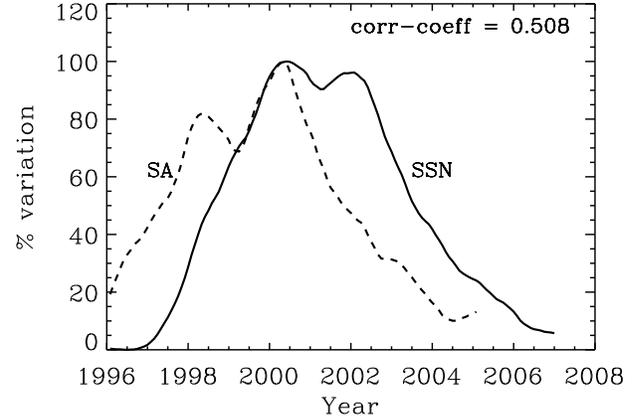}  
  \caption[]{\label{fig:hyst9606_ssa-ssn_shift} %
Sunspot area index (SA) is shifted to coincide its peak occurrence with that of the sunspot number (SSN) and the temporal offset is 23 months
 }
\end{figure}
Despite their similarity among many parameters during ascending and descending phases, their peak occurrences do not coincide well. From Fig.\,\ref{fig:hyst9606_ssa-ssn_shift}, it is quite evident that the correlations between two indicators drops heavily when the data is shifted in time to coincide their peak occurrences. For example the SSA show a correlation of 0.962 and it drops to 0.508  when their peaks are made to coincide.  Similarly the correlations of FI versus XBF (Not shown) and CME versus SSN (Not shown) drops to -0.356 and 0.642 respectively from their unshifted values of 0.862 and 0.874.   Therefore, considering the time delays based on the peak occurrences seems to be unreasonable and overall development of the time line of the solar cycle requires a greater attention. In order to bring out the detailed time lag information between the indicators, we perform cross-correlation analysis and the results are shown in Fig.3 wherein the cross-correlations of the indicators against sunspot area alone are shown.  Offsets (in months) of each considered parameter with respect to others at times of maximum correlations (99\% confidence) are given in Table 1. The offset times are mentioned in the form ''a/b'' where ''a'' is the time delay when the peak occurrences are considered as shown in Fig. 2 while ''b'' represent the time delay when the maximum of overall correlation is considered as shown in Fig. 3. An uncertainty of about a month can be seen in most of the computations performed here and can be attributed to the temporal resolution of the data used.

\subsection{Sunspot Number (SSN) and Area (SA)} 
In a strict sense SSN and SA time series are not supposed to show any temporal offset because  both of them are deduced from the same observed feature (sunspots) of the Sun. However,  during the cycle 21 SA is found lag behind the SSN by 23 months  (Fig. 1a) when maximum occurrences are considered and it reduced to 2 months (Figs. 1a \& 3a) when overall correlations are considered. Therefore, it appears that the delay between SA and SSN is an artificial one than real.  This behaviour though is not seen in cycle 22 (Fig. 1b \& 3b) is unambiguous in cycle 23 (Figs. 1c \& 3c) also. Earlier \citep{2008ApJ...686L..41R} have shown that the mismatch in the peak occurrences of SSN and XBF around solar maximum is due to the underestimations in SSN caused by the definition of SSN [SSN=k(10g+f), where SSN is sunspot number, k is proportionality constant, g is number of spot groups and f is the number of individual spots], wherein each spot group is assigned the same weight factor. \citet{2003SoPh..215..111T} also have explained the mismatch of major flare events with sunspot numbers in terms of such underestimations. A word of caution is therefore exercised here while using SSN as a standard time series to represent the solar activity particularly with reference to the maximum phase.

\subsection{10.7 cm solar  flux (F10)}
During solar cycle 23 the 10.7 cm radio flux peaks about twenty months later than the sunspot number while it leads sunspot area by two months (Table 3). Earlier studies indicated that this delay was  about one and half years even though the overall correlation was quiet strong at zero lag (Wilson et al., 1987).   During solar cycle 21 the F10 lags behind sunspot number by seventeen months when peak occurrences are considered while it is one month when overall correlations are considered.  However, during the cycle 22, F10 seem to be in phase with SSN \citep{2012JApA...33..387Y}.  Temporal offsets of F10 with sunspot area,  though is slightly higher when peak occurrences are considered, is only about a month when overall correlations are taken into account. Time lag identified with the peak occurrences of indicators may therefore, represent the local variability in the respective solar atmosphere from which these particular observed parameter originates. The common driving force may influence the source regions of different parameters in different ways that cause individual parameter peak to occur at different times although the long term trend in the variability may remain similar. Instrumental errors and data reduction methods may also influence the time of peak occurrences of independent parameters. We, therefore, opine that the time delays be considered only on overall basis than identifying them with respect to the peak occurrences.  Herein after the parameter “b” in table 1 is considered to represent temporal offsets for further analysis.
It is now well established that the compact,  sunspot-associated strong field regions are the dominant sources of S-component of 10.7 cm microwave radiation and that the nonthermal gyro-resonant emission is the contributing mechanism \citep{1982ApJ...255L.111L,chi82,1985ApJ...292..733S}.   However, it is to be noted that, while the magnetic field emerge to the surface, sunspots or dark pores can only form if the magnetic field strength exceeds about 1500 G \citep{2012ApJ...757L...8L}.  Also the pores or spots formed at opacity minimum region (OMR) around 50 km below the visible surface need not always appear at the visible surface of the Sun in the red continuum while they appear with little doubt in the near infrared continuum \citep{fou90}.  Therefore,  an emerging flux region that may have formed one or two rotations earlier and not encountered the threshold value to form sunspots but sufficient enough to trigger the microwave emission could be a cause of F10 leading sunspot time series. Similarly, there can be a delay of F10 with respect to sunspots in case of microwave emission continuing in the presence of remnant fields but insufficient to retain the stature of  sunspot. \citet{2006SoPh..234..393R}, in fact, have shown that the remnant fields remaining two rotations after the disappearance of sunspot can lead to soft x-ray emission in the spectral range 20-30 A.  \citet{bach04} have indicated that the solar cycle differences with temporal offsets for a pair of indices may be due to short-term phenomena \- possibly differences in timing and strengths of typical active regions during the course of each solar cycle.  It is therefore very important that co-temporal full disk multi-wavelength  observations on a continued basis be made in order to understand the temporal offsets with particular reference to cycle-to-cycle variations.  In particular, the observations in near infrared that can represent OMR, and G-band that represents well the regions of temperature minimum would help identifying unambiguously all the flux emerging regions (that include sunspots, pores, moving magnetic regions) to buildup a representative time series of the visible features of the Sun. 
\begin{table*}
    \centering
     \begin{tabular}{c c c c c c c}
     \hline\hline
$X\rightarrow$ & SSN & SA & F10.7 & CI & FI & XBF \\
$Y\downarrow$ \\
\hline
SA & -23/-2 \\
F10.7 & -17/-1 & 6/1 \\
CI & -26/-6 & -4/-5 & -8/-7 \\
FI & -13/-2 & 9/0 & 4/-2 & 13/6 \\
XBF & -18/-3 & 4/0 & -1/-2 & 9/6 & -4/-1 \\
CaI & -1/0 & 21/0 & 16/0 & 24/2 & 12/1 & 16/2 \\
\hline
\end{tabular}
\caption{ Temporal offsets (in months) as per Peak occurrences/Overall correlations of corresponding solar activity indicators for Solar cycle 21. Negative sign indicates Y lags X and Positive sign indicates Y leads X }
\end{table*}
\quad
\begin{table*}
\centering
\begin{tabular}{c c c c c c c c c}
\tableline\tableline
$X\rightarrow$ & SSN & SA & F10.7 & CI & FI & XBF & IRR & CaI\\
$Y\downarrow$ \\
\tableline
SA & 0/0 \\
F10.7 & 0/-1 & 0/-1 \\
CI & 0/0 & 0/1 & 0/1 \\
FI & 0/1 & 0/1 & 0/1 & 0/0 \\
XBF & 0/0 & 0/0 & 0/1 & 0/0 & 0/-1 \\
IRR & -2/-2 & -3/-2 & -2/1 & -2/-2 & -3/-2 & -3/-2 \\
CaI & -24/-3 & -24/-2 & -24/-2 & -24/-3 & -25/-3 & -25/-2 & -21/-1 \\
MgII & 0/-1 & -1/-1 & 0/-1 & 0/-2 & -2/-2 & -2/-1 & 2/0 & 23/1 \\
\tableline
\end{tabular}
\caption{ Same as Table 1 for Solar cycle 22 }
\end{table*}
;\quad
\begin{table*}
\centering
\begin{tabular}{c c c c c c c c c}
\tableline\tableline
$X\rightarrow$ & SSN & SA & F10.7 & CI & FI & XBF & IRR & MgII\\
$Y\downarrow$ \\
\tableline
SA & -23/-4 \\
F10.7 & -20/-2 & 2/2 \\
CI & -19/-2 & 4/2 & 1/0 \\
FI & 4/4 & 28/8 & 26/6 & 24/5 \\
XBF & -19/-2 & 2/2 & 0/0 & 0/0 & -25/-7 \\
IRR & -6/-2 & 17/7 & 14/4 & 13/4 & -10/-2 & 14/4 \\
MgII & -21/-6 & 0/-1 & -1/-3 & -2/-3 & -27/-10 & -2/-3 & -16/-8 \\
nCME & -24/-10 & -2/-5 & -4/-8 & -5/-8 & -29/-17 & -4/-8 & -19/-13 & -3/-4 \\
\tableline
\end{tabular}
\caption{ Same as Table 1 for Solar cycle 23 }
\end{table*}

\begin{figure}[hbtp]
  \centering
  \includegraphics[width=\columnwidth]{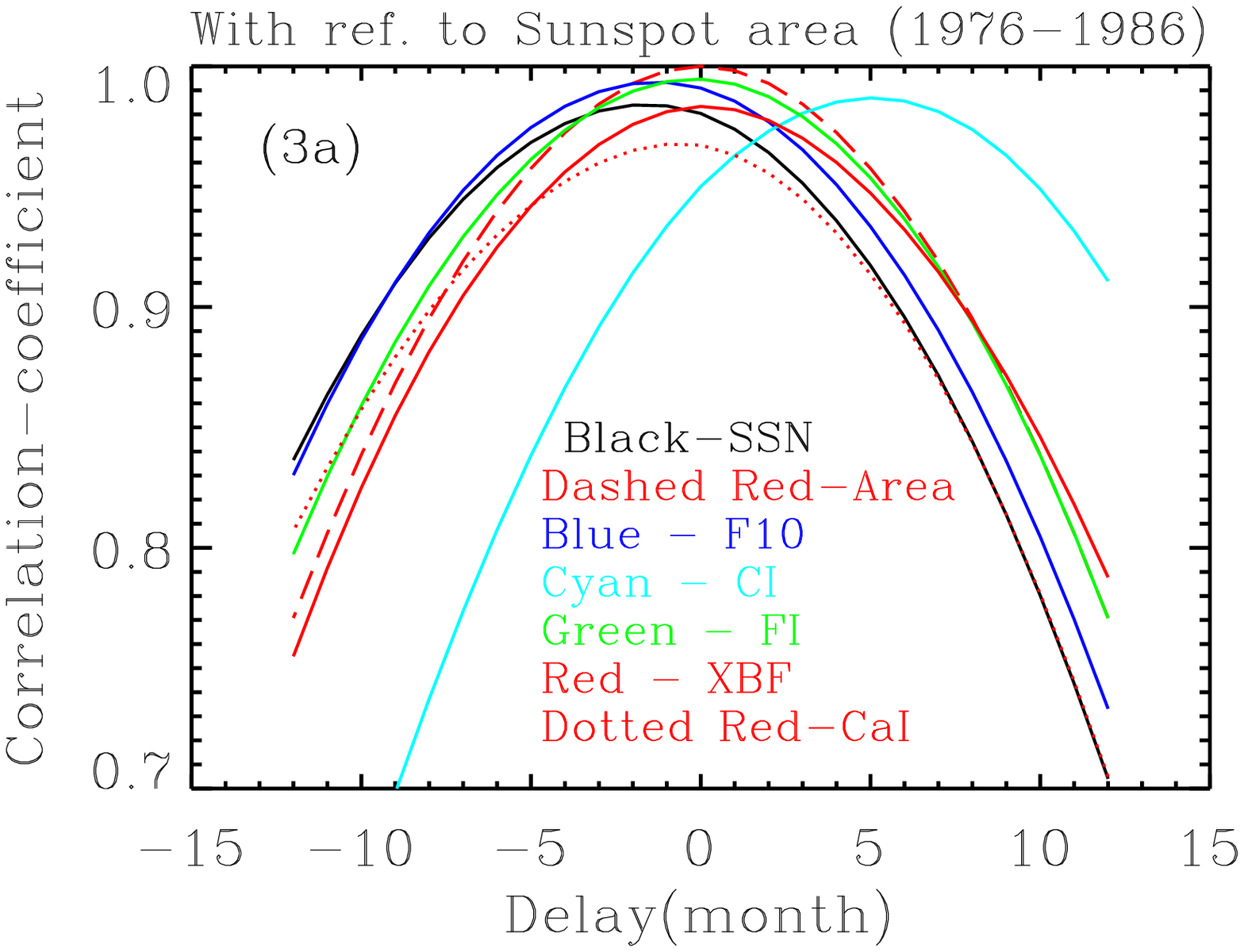}  
 \includegraphics[width=\columnwidth]{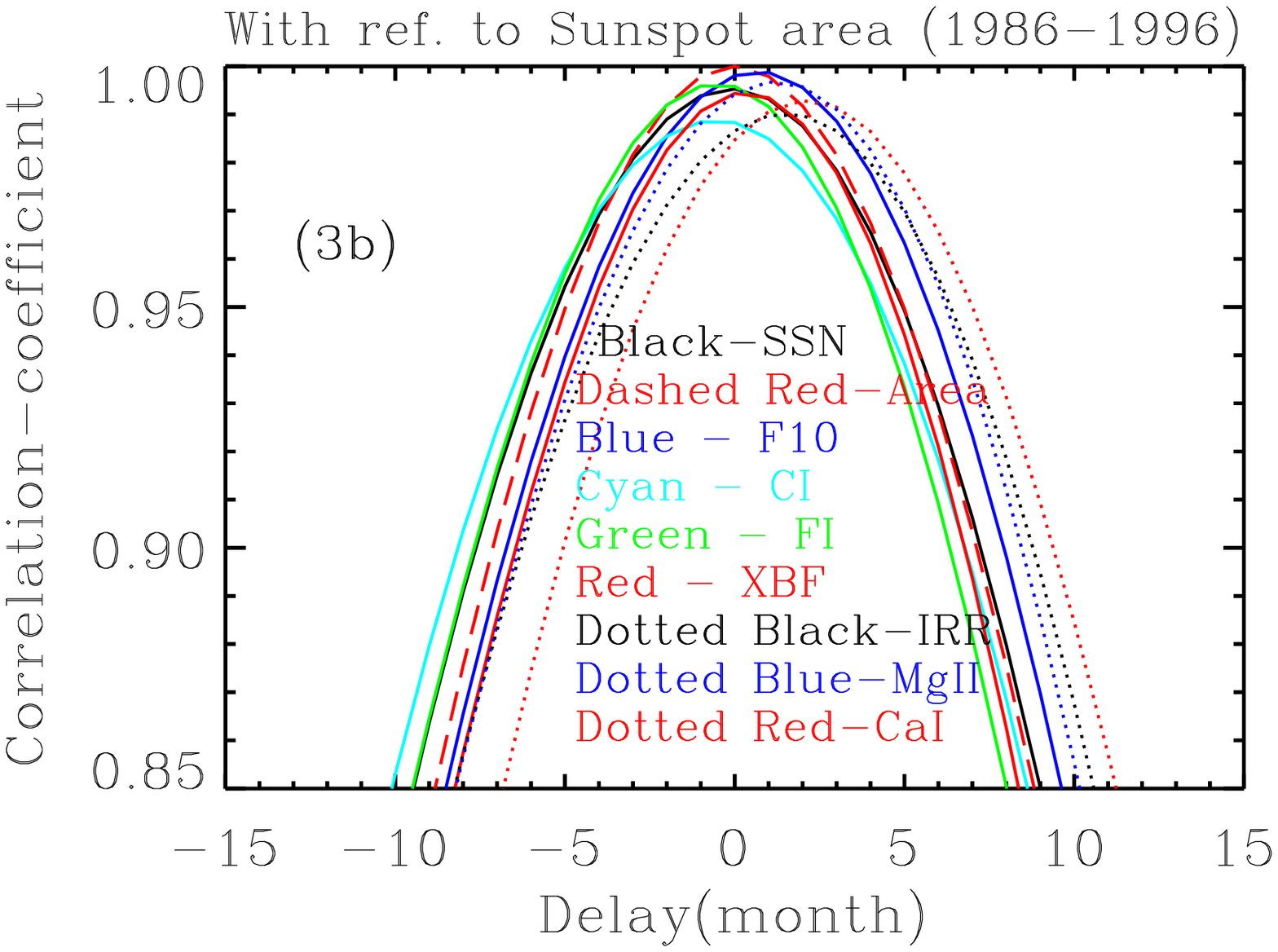} 
 \includegraphics[width=\columnwidth]{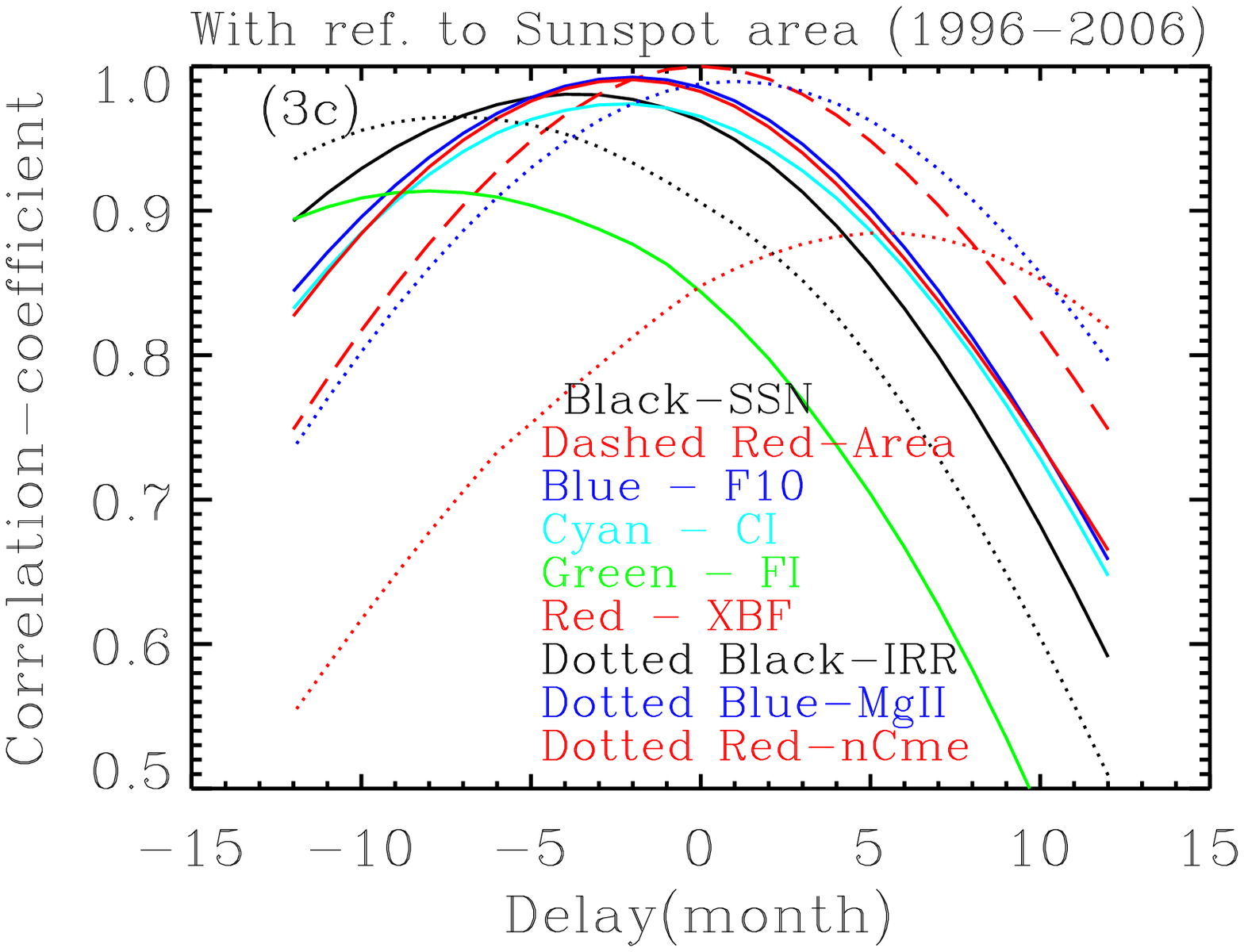}
  \caption[]{\label{fig:delay_wrt_ssa} %
3a, 3b and 3c depict the cross-correlations and time lags in months between sunspot area and other solar indices for solar cycles 21,22 and 23 respectively
 }
\end{figure}

\subsection{Coronal Index (CI)}

The coronal index (CI) of solar activity is the irradiance of the Sun as a star in the Fe XIV, coronal green line at mid visible wavelength of 5303 \AA. It is derived from ground-based observations of the green corona made by the network of coronal stations and brought to a common scale. \citet{2005JGRA..110.8106R} show that the Green coronal intensity and the sunspot number are highly correlated. Indeed in our analysis of CI with all the other indicators show high (99\% level) correlations (not shown). However, the temporal offsets are still seen in all the three solar cycles and are highly variable. In cycle 21 an offset of nearly five to six months is seen with all the parameters except with Ca plage index.  Similar is the situation in cycle 23 while in cycle 22 the temporal offsets are minimal.   It is interesting to see a temporal offset of about two to three months with Ca plage index and MgII in both cycles 21 and 23.   More interestingly CI leads the  chromospheric index MgII in both 22nd and 23rd cycles.The Coronal emission (CI) is known to be associated well with the plages which are in turn associated with sunspot regions (\citep{1999SoPh..188...99R}; \citep{1987PAICz..66...85R}) and that the emission corona is quite sensitive to the underlying photospheric field \citep{1997ApJ...485..419W}.  We therefore, opine that the fields emerged two to three rotations before they surface as spots with threshold strengths of about 1500 G \citep{2012ApJ...757L...8L} may give rise to the coronal emission  in advance of the spot or plage activity and hence may explain the delayed offsets.

\subsection{Mg II core to wing ratio (MgII) and Calcium Plage index (CaI)}
Mg II, the irradiance ratio from proximate wavelength ranges \citep{1999JGR...104.9995V} and relatively uncorrupted by instrumental effects,represents well the chromospheric activity. It originates at slightly higher in the chromosphere when compared to the regions where CaI originates. Though both MgII and CaI show clear double humped pattern,their peak occurrences do not match. During cycle 22, MgII show its maximum occurrence at the first peak of the double humped pattern while CaI show its maximum occurrence at the second peak (Fig. 1b). This difference brings in a time offset of 23 months (Table 2) when their peak occurrences are considered while the overall correlations show an offset of one month(within the limits of temporal resolution of the data) . Hence it is clear that peak occurrences indicate different manifestations of physical mechanisms responsible for their formation. During cycle 23 MgII show large deviations(nearly 1-10 months offsets) even when the overall correlations are considered.Although MgII  shows an offset of six months with SSN ,it does not show any offset with SA. The offsets with F10, CI and XBF are found to be minimal (around three months). It shows a delayed response of nearly 8 and 10 months with IRR and FI respectively. Since the originating regions of MgII (chromosphere) and IRR (photosphere) are different, their temporal offsets may be understood in terms of originating mechanisms.
Ca plage index does not show any temporal offset with sunspot activity (both SSN and SA) and F10 in cycle 21 (Table 1) that supports the results of \citep{bert10}.  However, it is difficult to reconcile the non-zero offsets of Ca plage index with SSN, SA and F10  in cycles 22 (Table 2).   Also it leads the coronal indices (CI, XBF) and the indices of energetic events,  FI in cycle 21 (Table 1) and lags behind them in cycle 22 (Table 2).   Understanding the non-zero offsets of Ca plage index with that of the spot related indices may be possible with the phenomenon of ''invisible sunspots'' (Fig 1 of \cite{fou90}.    Plages may emerge in the chromospheric heights even with comparatively lower strengths of the field in the previous one or two rotations while the spots may not have formed on the visible photospheric regions.   Similarly plages would remain even after the field strengths reduce below certain levels but continue to exist for few more rotations after the spot disappears.   The invisible sunspots \citep{fou90}, the field strengths of which have not crossed the threshold strengths of about 1500 G \citep{2012ApJ...757L...8L} may very sensitively influence the variability in the Ca index. In both these cases the plage index may appear to lead or lag the strong field sensitive indices like CI, XBF and FI.   We argue that the non-zero offsets of plage index with the coronal indices, F10, XBF and FI, in cycle 22 may be because of the different response times of chromosphere and coronal regions to the magnetic field. Perhaps the long-term average of short-term effects due to many active regions emerging and decaying during the course of the solar cycle play a role. However, it is pertinent to mention here that the lag or lead of an indicator with reference to the other may depend on the sensitivity of the region from which the index arises to the features beneath. 

\subsection{Flare-index (FI)}

FI, the flare index \citep{1952BAICz...3...52K} that is roughly proportional to the total energy emitted by the flare  \citep{2003SoPh..214..375O,2004SoPh..223..287O} seems to follow the SSN, SA, F10, CI and XBF well in the ascending and descending phases of the cycles 21 and 22 (Figs. 1a \& 1b) while during maximum phase of the cycle, deviations are clearly seen. However, the variability pattern during the cycle 23 show large deviations in both ascending and maximum phases of the cycle.  Overall delay of two months is seen with SSN and F10 \citep{2012JApA...33..387Y} during the cycle 21 but with SA it is near cotemporal.  FI displays no offset or offsets within the uncertainty of one month during cycle 22.  However, a considerable lead (Table 3) is seen with all the other parameters during the cycle 23 \citep{atac06}. 

Since the spotless flares are low energetic \citep{1982AcASn..23...95L} and that in our analysis the minimum value in the time series is subtracted out, we assume that the considered percentage variation of flare index represent only the active regions associated component. \citet{2012JApA...33..387Y} speculated that the lead of FI over sunspot activity during cycle 23 could be an indication that the ascending phase may contribute largely in terms of total energy released by flares. In the recent past it was argued that the active regions that undergo large changes in sunspot area are most flare productive \citep{chou13}.  We opine that the active regions evolving quite quickly to reach their threshold energy levels to give out flares are more in number during ascending and maximum phases of the cycle than in descending phase. Therefore,  the active regions evolving quite quickly to reach their threshold energy levels during ascending phase of the cycles, as in the case of solar cycle 23, might have caused more number of flares and in turn lead to peak occurrence of flare index had happened well in advance of the peak occurrence of SSN or SA.  However, similar situation may not arise in all the cycles because the evolving spots may not satisfy the other condition, namely,   the free energy content of the active region exceeding 50\% of the total energy \citep{chou13}.  It needs to be ascertained whether the evolution of flare productive active regions do really follow the phase of the solar cycle with particular reference to the growth phase.

\subsection{Background X-ray Flux (XBF)}

XBF, a coronal index of solar activity,  seems to show different temporal offsets with different time series in the three solar cycles considered here.   Different temporal offsets of XBF with SSN and SA has been discussed in terms of underestimations in SSN \citep{2008ApJ...686L..41R}.    XBF does not show any temporal offset with F10 in cycles 22 and 23 while lags behind it by two months in the cycle 21.  It is known that both XBF and F10 arise from coronal regions of closed magnetic flux loops. Since both the parameters are the emission measures of two spectral regions and that both of them do not represent the measures of energetic events such as flares, we believe that the observed lag, nearly same as the temporal resolution of the data used, may partially arise from statistical or instrumental effects. Magnetic complexity within active regions that varies from cycle to cycle, particularly near solar maximum may influence the individual physical mechanisms at work.  However, with a minimal lag in cycle 21 and no-offset condition in cycles 22 and 23 between XBF and F10 suggests that the s-components of them represent well the true strength of the solar cycle. A similar situation is seen with CI during the cycles 22 and 23, while a considerable lead (six months) is observed with it in cycle 21 (Table 1).  It is to be noted that the index CI, representing the total flux of energy emitted by the 5303 \AA coronal line, is derived from the limb observations \citep{1975BAICz..26..367R} and further extrapolated on to the disk facing the Earth.  Therefore, possible source of the temporal offset between these two coronal indices could be the inadequate contribution of evolutionary component of green line emission from the regions of corona in our line of sight to the daily values of CI.  This explanation however, is questionable in case of no-offset situation between XBF and CI in cycles 22 and 23. 

\subsection{Total Solar Irradiance (IRR)}

During solar cycle 22 the profile of IRR matches well with most of the time profiles of other parameters (Fig. 1b) and the offsets (one to two months) remain  within the uncertainties.  During cycle 23 IRR seems to deviate to a large extent with respect to the time profiles of other parameters (Fig. 1c) and the offsets are around two to thirteen months. IRR, the time  series mainly representing the radiation arising from the visible surface of the Sun, is known to depict the 11-year solar cycle variation (Fig. 1).  Although on shorter time scales IRR shows dips at times of passage of sunspots across the Sun's disk, the increase in IRR is well maintained by the features like faculae and network. All together these three-components still will not be able to represent completely the 0.1\% variability in IRR. In the recent past it was suggested that a fourth component,  a change of the global temperature of Sun modulated by the strength of activity,  can potentially  contribute to the long-term change of IRR \citep{fro13}. Such long term change in the global temperature of the Sun can influence the IRR differently in different solar cycles.    Cycle-to-cycle variation in the IRR may arise through the heat transport properties of the regions beneath the visible photosphere \citep{2004AdSpR..34..302K} and in turn complicate relationship of IRR with other indicators of solar activity, the evolutionary pattern of which may arise from the regions above the visible surface.  

\subsection{Occurrence rate of Coronal Mass Ejections (nCME)}

CMEs bring about large scale changes in the corona, which have fundamental implications for the evolution of the magnetic flux of the Sun. The rate of occurrence of CMEs  shows unambiguously the solar cycle related variability.  The rate of occurrence of CMEs was found to lag behind the sunspot numbers by about two years \citep{2003ESASP.535..403G} and with sunspot area by about three to five months \citep{2010ApJ...712L..77R}.  This discrepancy was interpreted as due to the underestimations of sunspot numbers and that the five months delay also was interpreted as a statistically insignificant \citep{2010ApJ...712L..77R}.   Present analysis show that the occurrence rate of CMEs lags behind not just SSN and SA but also F10, CI, FI, XBF, TSI and MgII although the lag time varies considerably (Table 3).  Minimum lag was found with MgII (four months) and maximum lag was found with another energetic event indicator, FI (seventeen months).  Magnetic field attaining the non-potentiality that represents the free energy content than the flux content as measured by the area of the active region, has been identified as a key parameter in producing CMEs.  nCME being an event representing indicator therefore may show temporal offsets with  F10, CI, XBF,  MgII which are broadly identified as the emission measures of the total energy content of the active regions.  nCME also shows large temporal offset with TSI that represents the global parameter of the Sun.   It is also known that the CMEs and flares do not occur on one-to-one basis \citep{2013SoPh..tmp..236N} and therefore a delay in their occurrence rate may be an expected one although which indicator leads the other remains unclear.  Also FI, represents the total energy emitted by the flare and hence is distinctly different from the nCME.    Therefore, in our opinion, the temporal offsets of nCME with other parameters do not come as a surprise and that comparing the events representing indicators such as nCME with others may lead to unreasonable inferences.

\section{Hysteresis pattern}

\begin{figure}[hbtp]
  \centering
  \includegraphics[width=\columnwidth]{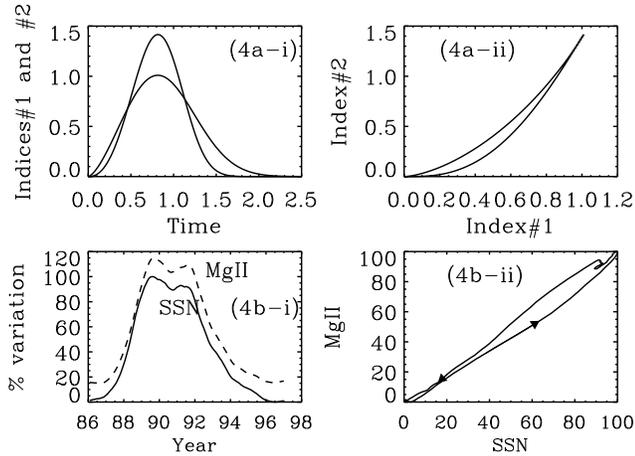}  
  \caption[]{\label{fig:dis8696_hysttest} %
4a-i shows two arbitrary indices \#\ 1 and \#\ 2  varying disproportionately with one another during ascending and descending phases while their peak occurrence coincide. 4a-ii  depicts a narrow loop pattern  when one index is plotted against another. Time evolution of annually smoothed sunspot number (SSN) and Magnesium core to wing ratio (MgII)  during solar cycle 22 are shown in Fig. 4b-i. MgII profile has been raised by about 15\%\ to avoid overlap of the curves. Hysteresis pattern of the indicators plotted in Fig. 4b-i is shown in Fig. 4b-ii
 }
\end{figure}

Any two indices with similar or proportionate rise and fall would result in a straight line in their scatter plot. Instead, they depict a loop like pattern if there is a temporal offset in their overall rise and fall [Fig. 1 of \citet{bach04}].   However, a loop like structure (Fig. 4a-ii) still can result in case of disproportionate rise and fall of the indices while their peak occurrence coincide (Fig. 4a-i). A typical example of such a case is the relationship between Smoothed SSN and the MgII (Fig. 4b-i) that results in a loop like pattern (Fig 4b-ii) even though the peaks of both the indices  occurred  at the same time.    Such variability, indicate the sensitivity of the chromosphere to the features underneath and would  provide better insight into the mechanisms through which the energy is transferred from one region of the solar atmosphere to the other.  However, the hysteresis pattern may not represent the direct cause and effect relationship because both of them may have a common source that triggers the originating mechanism at different regions of the solar atmosphere while their evolutionary pattern might vary significantly. Also the arbitrary, unphysical definition of sunspot number plays a role. It is to be noted that a single pattern of hysteresis is not maintained in the three solar cycles studied here (All the hysteresis plots are not shown here) and also the direction of hysteresis pattern (Lag or lead in terms of temporal offsets) between any two parameters is not consistent from one cycle to the other, although cycle to cycle variation, particularly between odd and even cycles were reported \citep{2003SoPh..215..111T,2005ICRC....2..139S}. However such differences needs to be verified for more number of odd and even cycles. In our analysis we find that in cycle 22, where the variation seem to be much less than the odd cycles 21 and 23, the hysteresis seem to exist eventhough the peaks of SSN and MgII coincide (Fig .4b). Also the limitations in the observations (such as the unseen sunspots in the visible continuum) may add to the varying hysteresis pattern. Therefore we opine and align with the conclusions of \citet{bach94} that the temporal offsets and in turn the hysteresis patterns among different solar indicators  seen here are mainly the consequences of the evolution of active regions from the subsurface to the coronal regions.   Also the sensitivity of the solar atmosphere to the evolving fields play an important role.   However, we believe that it is very hard to decide on the indicator (among the ones considered in this work) that truly represent the solar cycle. 
\section{Conclusions} 
Firstly the different methods of smoothing of time series may give rise to different delayed patterns among the solar cycle representing parameters. It is very important to smooth the time series very optimally without loosing the information at different time scales before comparing them for their delayed behaviour.
\begin{enumerate}
\item Since the sunspot number and area are derived from the same feature (sunspot) observed on the Sun, temporal offset is not expected between them.   The temporal offset seen between them therefore, is interpreted in terms of the underestimations of SSN using the relation SSN=k(10g+f) where a weight factor of ten is considered for each spot group.
\item Sunspot number and area both may have underestimations due to the unidentified spot groups (invisible sunspots)  that do not appear on visible photosphere while they can be quite active to influence the chromosphere and coronal regions and cause variability in their respective indicator time series.  These underestimations therefore may lead to offsets in the temporal profiles of the other activity indicators against them.
\item Considering peak occurrences of the time profiles of the solar activity indicators to identify the offsets may lead to unrealistic interpretations of the physical mechanisms responsible for their variability.
\item Evolutionary aspects of the time profiles of chromospheric and coronal indicators during ascending and descending phases of the solar cycle may cause offsets with respect to one another \citep{bach94}.  Such variability, particularly indicate the sensitivity of the upper atmosphere to the features underneath and would  provide better insight into the mechanisms through which the energy is transferred from one region of the solar atmosphere to the other. 
\item Continued observations of strong field regions of solar activity such as sunspots, pores, moving magnetic regions using sensitive detectors to unambiguously identify and record them to build up suitable representative time series are required to understand the true offsets among the solar indicators.
\item Also the sensitivity of the solar atmosphere to the evolving fields play an important role. However, we believe that it is very hard to decide on the indicator (among the ones considered in this work)  that truly represent the solar cycle.
\end{enumerate}

\acknowledgments
The authors acknowledge all the teams responsible for producing the elaborate data base of  all the solar indicators used in this work.  The sunspot-area data were complied by Royal Greenwich Observatory (RGO) and the Solar Optical Observing Network (SOON) of the US Air Force (USAF)/US National Oceanic and Atmospheric Administration (NOAA).  SOHO is a project of international cooperation between ESA and NASA. The LASCO/CME catalog is generated and maintained at the CDAW Data Center by NASA and The Catholic University of America in cooperation with the Naval Research Laboratory. The flare-index data used in this study were calculated by T. Ata{\c c} and A. {\"O}zg{\"u}{\c c} from Bogazici University’s Kandilli Observatory, Istanbul, Turkey.

\noindent
\nocite{*}
 \bibliographystyle{spr-mp-nameyear-cnd}  
 \bibliography{biblio}                

\begin{thebibliography}{48}
\ifx \bisbn   \undefined \def \bisbn  #1{ISBN #1}\fi
\ifx \binits  \undefined \def \binits#1{#1} \fi
\ifx \bauthor  \undefined \def \bauthor#1{#1} \fi
\ifx \batitle  \undefined \def \batitle#1{#1} \fi
\ifx \bjtitle  \undefined \def \bjtitle#1{#1}\fi
\ifx \bvolume  \undefined \def \bvolume#1{\textbf{#1}}\fi
\ifx \byear  \undefined \def \byear#1{#1} \fi
\ifx \bissue  \undefined \def \bissue#1{#1} \fi
\ifx \bfpage  \undefined \def \bfpage#1{#1} \fi
\ifx \blpage  \undefined \def \blpage #1{#1} \fi
\ifx \burl  \undefined \def \burl#1{\textsf{#1}} \fi
\ifx \doiurl  \undefined \def \doiurl#1{\textsf{#1}} \fi
\ifx \betal  \undefined \def \betal{\textit{et al.}} \fi
\ifx \binstitute  \undefined \def \binstitute#1{#1} \fi
\ifx \binstitutionaled  \undefined \def \binstitutionaled#1{#1} \fi
\ifx \bctitle  \undefined \def \bctitle#1{#1} \fi
\ifx \beditor  \undefined \def \beditor#1{#1} \fi
\ifx \bpublisher  \undefined \def \bpublisher#1{#1} \fi
\ifx \bbtitle  \undefined \def \bbtitle#1{#1} \fi
\ifx \bedition  \undefined \def \bedition#1{#1} \fi
\ifx \bseriesno  \undefined \def \bseriesno#1{#1} \fi
\ifx \blocation  \undefined \def \blocation#1{#1} \fi
\ifx \bsertitle  \undefined \def \bsertitle#1{#1} \fi
\ifx \bsnm \undefined \def \bsnm#1{#1} \fi
\ifx \bsuffix \undefined \def \bsuffix#1{#1} \fi
\ifx \bparticle \undefined \def \bparticle#1{#1} \fi
\ifx \barticle \undefined \def \barticle#1{#1} \fi
\ifx \bconfdate \undefined \def \bconfdate #1{#1} \fi
\ifx \botherref \undefined \def \botherref #1{#1} \fi
\ifx \url \undefined \def \url#1{\textsf{#1}} \fi
\ifx \bchapter \undefined \def \bchapter#1{#1} \fi
\ifx \bbook \undefined \def \bbook#1{#1} \fi
\ifx \bcomment \undefined \def \bcomment#1{#1} \fi
\ifx \oauthor \undefined \def \oauthor#1{#1} \fi
\ifx \citeauthoryear \undefined \def \citeauthoryear#1{#1} \fi
\ifx \endbibitem  \undefined \def \endbibitem {}\fi
\ifx \bconflocation  \undefined \def \bconflocation#1{#1} \fi
\ifx \arxivurl  \undefined \def \arxivurl#1{\textsf{#1}} \fi

\bibitem[\protect\citeauthoryear{{Ata{\c c}} and {{\"O}zg{\"u}{\c
  c}}}{2006}]{atac06}
\begin{barticle}
\bauthor{\bsnm{{Ata{\c c}}}, \binits{T.}},
\bauthor{\bsnm{{{\"O}zg{\"u}{\c c}}}, \binits{A.}}:
\bjtitle{\solphys}
\bvolume{233},
\bfpage{139}
(\byear{2006})
\end{barticle}
\endbibitem

\bibitem[\protect\citeauthoryear{{Bachmann} and {White}}{1994}]{bach94}
\begin{barticle}
\bauthor{\bsnm{{Bachmann}}, \binits{K.T.}},
\bauthor{\bsnm{{White}}, \binits{O.R.}}:
\bjtitle{\solphys}
\bvolume{150},
\bfpage{347}
(\byear{1994})
\end{barticle}
\endbibitem

\bibitem[\protect\citeauthoryear{{Bachmann} et~al.}{2004}]{bach04}
\begin{barticle}
\bauthor{\bsnm{{Bachmann}}, \binits{K.T.}},
\bauthor{\bsnm{{Maymani}}, \binits{H.}},
\bauthor{\bsnm{{Nautiyal}}, \binits{K.}},
\bauthor{\bsnm{{te Velde}}, \binits{V.}}:
\bjtitle{Advances in Space Research}
\bvolume{34},
\bfpage{274}
(\byear{2004})
\end{barticle}
\endbibitem

\bibitem[\protect\citeauthoryear{{Bai}}{1993}]{1993ApJ...404..805B}
\begin{barticle}
\bauthor{\bsnm{{Bai}}, \binits{T.}}:
\bjtitle{\apj}
\bvolume{404},
\bfpage{805}
(\byear{1993})
\end{barticle}
\endbibitem

\bibitem[\protect\citeauthoryear{{Bertello} et~al.}{2010}]{bert10}
\begin{barticle}
\bauthor{\bsnm{{Bertello}}, \binits{L.}},
\bauthor{\bsnm{{Ulrich}}, \binits{R.K.}},
\bauthor{\bsnm{{Boyden}}, \binits{J.E.}}:
\bjtitle{\solphys}
\bvolume{264},
\bfpage{31}
(\byear{2010})
\end{barticle}
\endbibitem

\bibitem[\protect\citeauthoryear{{Bromund} et~al.}{1995}]{bro95}
\begin{barticle}
\bauthor{\bsnm{{Bromund}}, \binits{K.R.}},
\bauthor{\bsnm{{McTiernan}}, \binits{J.M.}},
\bauthor{\bsnm{{Kane}}, \binits{S.R.}}:
\bjtitle{\apj}
\bvolume{455},
\bfpage{733}
(\byear{1995})
\end{barticle}
\endbibitem

\bibitem[\protect\citeauthoryear{{Charbonneau}}{2010}]{char10}
\begin{barticle}
\bauthor{\bsnm{{Charbonneau}}, \binits{P.}}:
\bjtitle{Living Rev. Solar Phys.}
\bvolume{7},
\bfpage{3}
(\byear{2010})
\end{barticle}
\endbibitem

\bibitem[\protect\citeauthoryear{{Chiuderi-Drago} et~al.}{1982}]{chi82}
\begin{barticle}
\bauthor{\bsnm{{Chiuderi-Drago}}, \binits{F.}},
\bauthor{\bsnm{{Bandiera}}, \binits{R.}},
\bauthor{\bsnm{{Willson}}, \binits{R.F.}},
\bauthor{\bsnm{{Slottje}}, \binits{C.}},
\bauthor{\bsnm{{Falciani}}, \binits{R.}},
\bauthor{\bsnm{{Antonucci}}, \binits{E.}},
\bauthor{\bsnm{{Lang}}, \binits{K.R.}},
\bauthor{\bsnm{{Shibasaki}}, \binits{K.}}:
\bjtitle{\solphys}
\bvolume{80},
\bfpage{71}
(\byear{1982})
\end{barticle}
\endbibitem

\bibitem[\protect\citeauthoryear{{Choudhary} et~al.}{2013}]{chou13}
\begin{barticle}
\bauthor{\bsnm{{Choudhary}}, \binits{D.P.}},
\bauthor{\bsnm{{Gosain}}, \binits{S.}},
\bauthor{\bsnm{{Gopalswamy}}, \binits{N.}},
\bauthor{\bsnm{{Manoharan}}, \binits{P.K.}},
\bauthor{\bsnm{{Chandra}}, \binits{R.}},
\bauthor{\bsnm{{Uddin}}, \binits{W.}},
\bauthor{\bsnm{{Srivastava}}, \binits{A.K.}},
\bauthor{\bsnm{{Yashiro}}, \binits{S.}},
\bauthor{\bsnm{{Joshi}}, \binits{N.C.}},
\bauthor{\bsnm{{Kayshap}}, \binits{P.}},
\bauthor{\bsnm{{Dwivedi}}, \binits{V.C.}},
\bauthor{\bsnm{{Mahalakshmi}}, \binits{K.}},
\bauthor{\bsnm{{Elamathi}}, \binits{E.}},
\bauthor{\bsnm{{Norris}}, \binits{M.}},
\bauthor{\bsnm{{Awasthi}}, \binits{A.K.}},
\bauthor{\bsnm{{Jain}}, \binits{R.}}:
\bjtitle{Advances in Space Research}
\bvolume{52},
\bfpage{1561}
(\byear{2013})
\end{barticle}
\endbibitem

\bibitem[\protect\citeauthoryear{{Donnelly}}{1991}]{don91}
\begin{barticle}
\bauthor{\bsnm{{Donnelly}}, \binits{R.F.}}:
\bjtitle{J.Geomagn.Geoelectr.Suppl}
\bvolume{43},
\bfpage{835}
(\byear{1991})
\end{barticle}
\endbibitem

\bibitem[\protect\citeauthoryear{{Dorman} et~al.}{1999}]{dor99}
\begin{bchapter}
\bauthor{\bsnm{{Dorman}}, \binits{L.I.}},
\bauthor{\bsnm{{Villoresi}}, \binits{G.}},
\bauthor{\bsnm{{Dorman}}, \binits{I.V.}},
\bauthor{\bsnm{{Iucci}}, \binits{N.}},
\bauthor{\bsnm{{Parisi}}, \binits{M.}}:
In: \beditor{\bsnm{{Suess}}, \binits{S.T.}},
\beditor{\bsnm{{Gary}}, \binits{G.A.}},
\beditor{\bsnm{{Nerney}}, \binits{S.F.}} (eds.)
\bbtitle{American Institute of Physics Conference Series}.
\bsertitle{American Institute of Physics Conference Series},
vol. \bseriesno{471},
p. \bfpage{621}
(\byear{1999})
\end{bchapter}
\endbibitem

\bibitem[\protect\citeauthoryear{{Foukal} et~al.}{1990}]{fou90}
\begin{barticle}
\bauthor{\bsnm{{Foukal}}, \binits{P.}},
\bauthor{\bsnm{{Little}}, \binits{R.}},
\bauthor{\bsnm{{Graves}}, \binits{J.}},
\bauthor{\bsnm{{Rabin}}, \binits{D.}},
\bauthor{\bsnm{{Lynch}}, \binits{D.}}:
\bjtitle{\apj}
\bvolume{353},
\bfpage{712}
(\byear{1990})
\end{barticle}
\endbibitem

\bibitem[\protect\citeauthoryear{{Fr{\"o}hlich}}{2013}]{fro13}
\begin{barticle}
\bauthor{\bsnm{{Fr{\"o}hlich}}, \binits{C.}}:
\bjtitle{\ssr}
\bvolume{176},
\bfpage{237}
(\byear{2013})
\end{barticle}
\endbibitem

\bibitem[\protect\citeauthoryear{{Gopalswamy}
  et~al.}{2003}]{2003ESASP.535..403G}
\begin{bchapter}
\bauthor{\bsnm{{Gopalswamy}}, \binits{N.}},
\bauthor{\bsnm{{Lara}}, \binits{A.}},
\bauthor{\bsnm{{Yashiro}}, \binits{S.}},
\bauthor{\bsnm{{Nunes}}, \binits{S.}},
\bauthor{\bsnm{{Howard}}, \binits{R.A.}}:
In: \beditor{\bsnm{{Wilson}}, \binits{A.}} (ed.)
\bbtitle{Solar Variability as an Input to the Earth's Environment}.
\bsertitle{ESA Special Publication},
vol. \bseriesno{535},
p. \bfpage{403}
(\byear{2003})
\end{bchapter}
\endbibitem

\bibitem[\protect\citeauthoryear{{Jimenez-Reyes} et~al.}{1998}]{jim98}
\begin{barticle}
\bauthor{\bsnm{{Jimenez-Reyes}}, \binits{S.J.}},
\bauthor{\bsnm{{Regulo}}, \binits{C.}},
\bauthor{\bsnm{{Palle}}, \binits{P.L.}},
\bauthor{\bsnm{{Roca Cortes}}, \binits{T.}}:
\bjtitle{\aap}
\bvolume{329},
\bfpage{1119}
(\byear{1998})
\end{barticle}
\endbibitem

\bibitem[\protect\citeauthoryear{{Kane}}{2002}]{2002AnGeo..20..741K}
\begin{barticle}
\bauthor{\bsnm{{Kane}}, \binits{R.P.}}:
\bjtitle{Annales Geophysicae}
\bvolume{20},
\bfpage{741}
(\byear{2002})
\end{barticle}
\endbibitem

\bibitem[\protect\citeauthoryear{{Kane}}{2003}]{2003JGRA..108.1379K}
\begin{barticle}
\bauthor{\bsnm{{Kane}}, \binits{R.P.}}:
\bjtitle{Journal of Geophysical Research (Space Physics)}
\bvolume{108},
\bfpage{1379}
(\byear{2003})
\end{barticle}
\endbibitem

\bibitem[\protect\citeauthoryear{{Kane}}{2011}]{2011SoPh..269..451K}
\begin{barticle}
\bauthor{\bsnm{{Kane}}, \binits{R.P.}}:
\bjtitle{\solphys}
\bvolume{269},
\bfpage{451}
(\byear{2011})
\end{barticle}
\endbibitem

\bibitem[\protect\citeauthoryear{{Kleczek}}{1952}]{1952BAICz...3...52K}
\begin{barticle}
\bauthor{\bsnm{{Kleczek}}, \binits{J.}}:
\bjtitle{Bulletin of the Astronomical Institutes of Czechoslovakia}
\bvolume{3},
\bfpage{52}
(\byear{1952})
\end{barticle}
\endbibitem

\bibitem[\protect\citeauthoryear{{Kuhn}}{2004}]{2004AdSpR..34..302K}
\begin{barticle}
\bauthor{\bsnm{{Kuhn}}, \binits{J.R.}}:
\bjtitle{Advances in Space Research}
\bvolume{34},
\bfpage{302}
(\byear{2004})
\end{barticle}
\endbibitem

\bibitem[\protect\citeauthoryear{{Lang} and
  {Willson}}{1982}]{1982ApJ...255L.111L}
\begin{barticle}
\bauthor{\bsnm{{Lang}}, \binits{K.R.}},
\bauthor{\bsnm{{Willson}}, \binits{R.F.}}:
\bjtitle{\apjl}
\bvolume{255},
\bfpage{111}
(\byear{1982})
\end{barticle}
\endbibitem

\bibitem[\protect\citeauthoryear{{Livingston}
  et~al.}{2012}]{2012ApJ...757L...8L}
\begin{barticle}
\bauthor{\bsnm{{Livingston}}, \binits{W.}},
\bauthor{\bsnm{{Penn}}, \binits{M.J.}},
\bauthor{\bsnm{{Svalgaard}}, \binits{L.}}:
\bjtitle{\apjl}
\bvolume{757},
\bfpage{8}
(\byear{2012})
\end{barticle}
\endbibitem

\bibitem[\protect\citeauthoryear{{Luo}}{1982}]{1982AcASn..23...95L}
\begin{barticle}
\bauthor{\bsnm{{Luo}}, \binits{B.-R.}}:
\bjtitle{Acta Astronomica Sinica}
\bvolume{23},
\bfpage{95}
(\byear{1982})
\end{barticle}
\endbibitem

\bibitem[\protect\citeauthoryear{{Nitta} et~al.}{2013}]{2013SoPh..tmp..236N}
\begin{botherref}
\oauthor{\bsnm{{Nitta}}, \binits{N.V.}},
\oauthor{\bsnm{{Aschwanden}}, \binits{M.J.}},
\oauthor{\bsnm{{Freeland}}, \binits{S.L.}},
\oauthor{\bsnm{{Lemen}}, \binits{J.R.}},
\oauthor{\bsnm{{W{\"u}lser}}, \binits{J.-P.}},
\oauthor{\bsnm{{Zarro}}, \binits{D.M.}}:
\solphys
(2013).
\arxivurl{1308.1465}
\end{botherref}
\endbibitem

\bibitem[\protect\citeauthoryear{{Ozguc} et~al.}{2002}]{2002cosp...34E.513O}
\begin{bchapter}
\bauthor{\bsnm{{Ozguc}}, \binits{A.}},
\bauthor{\bsnm{{Antalova}}, \binits{A.}},
\bauthor{\bsnm{{Atac}}, \binits{T.}}:
In: \bbtitle{34th COSPAR Scientific Assembly}.
\bsertitle{COSPAR Meeting},
vol. \bseriesno{34},
\byear{2002}
\end{bchapter}
\endbibitem

\bibitem[\protect\citeauthoryear{{{\"O}zg{\"u}{\c c}}
  et~al.}{2003}]{2003SoPh..214..375O}
\begin{barticle}
\bauthor{\bsnm{{{\"O}zg{\"u}{\c c}}}, \binits{A.}},
\bauthor{\bsnm{{Ata{\c c}}}, \binits{T.}},
\bauthor{\bsnm{{Ryb{\'a}k}}, \binits{J.}}:
\bjtitle{\solphys}
\bvolume{214},
\bfpage{375}
(\byear{2003})
\end{barticle}
\endbibitem

\bibitem[\protect\citeauthoryear{{{\"O}zg{\"u}{\c c}}
  et~al.}{2004}]{2004SoPh..223..287O}
\begin{barticle}
\bauthor{\bsnm{{{\"O}zg{\"u}{\c c}}}, \binits{A.}},
\bauthor{\bsnm{{Ata{\c c}}}, \binits{T.}},
\bauthor{\bsnm{{Ryb{\'a}k}}, \binits{J.}}:
\bjtitle{\solphys}
\bvolume{223},
\bfpage{287}
(\byear{2004})
\end{barticle}
\endbibitem

\bibitem[\protect\citeauthoryear{{{\"O}zg{\"u}{\c c}}
  et~al.}{2012}]{2012SoPh..281..839O}
\begin{barticle}
\bauthor{\bsnm{{{\"O}zg{\"u}{\c c}}}, \binits{A.}},
\bauthor{\bsnm{{Kilcik}}, \binits{A.}},
\bauthor{\bsnm{{Rozelot}}, \binits{J.P.}}:
\bjtitle{\solphys}
\bvolume{281},
\bfpage{839}
(\byear{2012})
\end{barticle}
\endbibitem

\bibitem[\protect\citeauthoryear{{Ramesh}}{2010}]{2010ApJ...712L..77R}
\begin{barticle}
\bauthor{\bsnm{{Ramesh}}, \binits{K.B.}}:
\bjtitle{\apjl}
\bvolume{712},
\bfpage{77}
(\byear{2010})
\end{barticle}
\endbibitem

\bibitem[\protect\citeauthoryear{{Ramesh} and
  {Lakshmi}}{2012}]{2012SoPh..276..395R}
\begin{barticle}
\bauthor{\bsnm{{Ramesh}}, \binits{K.B.}},
\bauthor{\bsnm{{Lakshmi}}, \binits{N.B.}}:
\bjtitle{\solphys}
\bvolume{276},
\bfpage{395}
(\byear{2012}).
\arxivurl{1109.2700}
\end{barticle}
\endbibitem

\bibitem[\protect\citeauthoryear{{Ramesh} and
  {Raman}}{2006}]{2006SoPh..234..393R}
\begin{barticle}
\bauthor{\bsnm{{Ramesh}}, \binits{K.B.}},
\bauthor{\bsnm{{Raman}}, \binits{K.S.}}:
\bjtitle{\solphys}
\bvolume{234},
\bfpage{393}
(\byear{2006})
\end{barticle}
\endbibitem

\bibitem[\protect\citeauthoryear{{Ramesh} and
  {Rohini}}{2008}]{2008ApJ...686L..41R}
\begin{barticle}
\bauthor{\bsnm{{Ramesh}}, \binits{K.B.}},
\bauthor{\bsnm{{Rohini}}, \binits{V.S.}}:
\bjtitle{\apjl}
\bvolume{686},
\bfpage{41}
(\byear{2008})
\end{barticle}
\endbibitem

\bibitem[\protect\citeauthoryear{{Ramesh} et~al.}{1999}]{1999SoPh..188...99R}
\begin{barticle}
\bauthor{\bsnm{{Ramesh}}, \binits{K.B.}},
\bauthor{\bsnm{{Nagabhushana}}, \binits{B.S.}},
\bauthor{\bsnm{{Varghese}}, \binits{B.A.}}:
\bjtitle{\solphys}
\bvolume{188},
\bfpage{99}
(\byear{1999})
\end{barticle}
\endbibitem

\bibitem[\protect\citeauthoryear{{Rybansky}}{1975}]{1975BAICz..26..367R}
\begin{barticle}
\bauthor{\bsnm{{Rybansky}}, \binits{M.}}:
\bjtitle{Bulletin of the Astronomical Institutes of Czechoslovakia}
\bvolume{26},
\bfpage{367}
(\byear{1975})
\end{barticle}
\endbibitem

\bibitem[\protect\citeauthoryear{{Rybansk{\'y}}
  et~al.}{1987}]{1987PAICz..66...85R}
\begin{barticle}
\bauthor{\bsnm{{Rybansk{\'y}}}, \binits{M.}},
\bauthor{\bsnm{{Ru{\v s}in}}, \binits{V.}},
\bauthor{\bsnm{{Dzif{\v c}{\'a}kov{\'a}}}, \binits{E.}}:
\bjtitle{Publications of the Astronomical Institute of the Czechoslovak Academy
  of Sciences}
\bvolume{66},
\bfpage{85}
(\byear{1987})
\end{barticle}
\endbibitem

\bibitem[\protect\citeauthoryear{{Rybansk{\'y}}
  et~al.}{2005}]{2005JGRA..110.8106R}
\begin{barticle}
\bauthor{\bsnm{{Rybansk{\'y}}}, \binits{M.}},
\bauthor{\bsnm{{Ru{\v s}in}}, \binits{V.}},
\bauthor{\bsnm{{Minarovjech}}, \binits{M.}},
\bauthor{\bsnm{{Klocok}}, \binits{L.}},
\bauthor{\bsnm{{Cliver}}, \binits{E.W.}}:
\bjtitle{Journal of Geophysical Research (Space Physics)}
\bvolume{110},
\bfpage{8106}
(\byear{2005})
\end{barticle}
\endbibitem

\bibitem[\protect\citeauthoryear{{Shevgaonkar} and
  {Kundu}}{1985}]{1985ApJ...292..733S}
\begin{barticle}
\bauthor{\bsnm{{Shevgaonkar}}, \binits{R.K.}},
\bauthor{\bsnm{{Kundu}}, \binits{M.R.}}:
\bjtitle{\apj}
\bvolume{292},
\bfpage{733}
(\byear{1985})
\end{barticle}
\endbibitem

\bibitem[\protect\citeauthoryear{{Singh} et~al.}{2005}]{2005ICRC....2..139S}
\begin{barticle}
\bauthor{\bsnm{{Singh}}, \binits{M.}},
\bauthor{\bsnm{{Badruddin}}},
\bauthor{\bsnm{{Ananth}}, \binits{A.G.}}:
\bjtitle{International Cosmic Ray Conference}
\bvolume{2},
\bfpage{139}
(\byear{2005})
\end{barticle}
\endbibitem

\bibitem[\protect\citeauthoryear{{Skumanich}
  et~al.}{1984}]{1984ApJ...282..776S}
\begin{barticle}
\bauthor{\bsnm{{Skumanich}}, \binits{A.}},
\bauthor{\bsnm{{Lean}}, \binits{J.L.}},
\bauthor{\bsnm{{Livingston}}, \binits{W.C.}},
\bauthor{\bsnm{{White}}, \binits{O.R.}}:
\bjtitle{\apj}
\bvolume{282},
\bfpage{776}
(\byear{1984})
\end{barticle}
\endbibitem

\bibitem[\protect\citeauthoryear{{Suyal} et~al.}{2012}]{2012SoPh..276..407S}
\begin{barticle}
\bauthor{\bsnm{{Suyal}}, \binits{V.}},
\bauthor{\bsnm{{Prasad}}, \binits{A.}},
\bauthor{\bsnm{{Singh}}, \binits{H.P.}}:
\bjtitle{\solphys}
\bvolume{276},
\bfpage{407}
(\byear{2012}).
\arxivurl{1112.5236}
\end{barticle}
\endbibitem

\bibitem[\protect\citeauthoryear{{Tapping} and
  {Detracey}}{1990}]{1990SoPh..127..321T}
\begin{barticle}
\bauthor{\bsnm{{Tapping}}, \binits{K.F.}},
\bauthor{\bsnm{{Detracey}}, \binits{B.}}:
\bjtitle{\solphys}
\bvolume{127},
\bfpage{321}
(\byear{1990})
\end{barticle}
\endbibitem

\bibitem[\protect\citeauthoryear{{Temmer} et~al.}{2003}]{2003SoPh..215..111T}
\begin{barticle}
\bauthor{\bsnm{{Temmer}}, \binits{M.}},
\bauthor{\bsnm{{Veronig}}, \binits{A.}},
\bauthor{\bsnm{{Hanslmeier}}, \binits{A.}}:
\bjtitle{\solphys}
\bvolume{215},
\bfpage{111}
(\byear{2003})
\end{barticle}
\endbibitem

\bibitem[\protect\citeauthoryear{{Viereck} and
  {Puga}}{1999}]{1999JGR...104.9995V}
\begin{barticle}
\bauthor{\bsnm{{Viereck}}, \binits{R.A.}},
\bauthor{\bsnm{{Puga}}, \binits{L.C.}}:
\bjtitle{\jgr}
\bvolume{104},
\bfpage{9995}
(\byear{1999})
\end{barticle}
\endbibitem

\bibitem[\protect\citeauthoryear{{Wagner}}{1988}]{1988AdSpR...8...67W}
\begin{barticle}
\bauthor{\bsnm{{Wagner}}, \binits{W.J.}}:
\bjtitle{Advances in Space Research}
\bvolume{8},
\bfpage{67}
(\byear{1988})
\end{barticle}
\endbibitem

\bibitem[\protect\citeauthoryear{{Wang} et~al.}{1997}]{1997ApJ...485..419W}
\begin{barticle}
\bauthor{\bsnm{{Wang}}, \binits{Y.-M.}},
\bauthor{\bsnm{{Sheeley}}, \binits{N.R.} \bsuffix{Jr.}},
\bauthor{\bsnm{{Hawley}}, \binits{S.H.}},
\bauthor{\bsnm{{Kraemer}}, \binits{J.R.}},
\bauthor{\bsnm{{Brueckner}}, \binits{G.E.}},
\bauthor{\bsnm{{Howard}}, \binits{R.A.}},
\bauthor{\bsnm{{Korendyke}}, \binits{C.M.}},
\bauthor{\bsnm{{Michels}}, \binits{D.J.}},
\bauthor{\bsnm{{Moulton}}, \binits{N.E.}},
\bauthor{\bsnm{{Socker}}, \binits{D.G.}},
\bauthor{\bsnm{{Schwenn}}, \binits{R.}}:
\bjtitle{\apj}
\bvolume{485},
\bfpage{419}
(\byear{1997})
\end{barticle}
\endbibitem

\bibitem[\protect\citeauthoryear{{Wheatland} and
  {Litvinenko}}{2001}]{2001ApJ...557..332W}
\begin{barticle}
\bauthor{\bsnm{{Wheatland}}, \binits{M.S.}},
\bauthor{\bsnm{{Litvinenko}}, \binits{Y.E.}}:
\bjtitle{\apj}
\bvolume{557},
\bfpage{332}
(\byear{2001})
\end{barticle}
\endbibitem

\bibitem[\protect\citeauthoryear{{Wilson} et~al.}{1987}]{1987SoPh..111..279W}
\begin{barticle}
\bauthor{\bsnm{{Wilson}}, \binits{R.M.}},
\bauthor{\bsnm{{Moore}}, \binits{R.L.}},
\bauthor{\bsnm{{Rabin}}, \binits{D.}}:
\bjtitle{\solphys}
\bvolume{111},
\bfpage{279}
(\byear{1987})
\end{barticle}
\endbibitem

\bibitem[\protect\citeauthoryear{{Yan} et~al.}{2012}]{2012JApA...33..387Y}
\begin{barticle}
\bauthor{\bsnm{{Yan}}, \binits{X.L.}},
\bauthor{\bsnm{{Deng}}, \binits{L.H.}},
\bauthor{\bsnm{{Qu}}, \binits{Z.Q.}},
\bauthor{\bsnm{{Xu}}, \binits{C.L.}},
\bauthor{\bsnm{{Kong}}, \binits{D.F.}}:
\bjtitle{Journal of Astrophysics and Astronomy}
\bvolume{33},
\bfpage{387}
(\byear{2012})
\end{barticle}
\endbibitem

\end{thebibliography}

\end{document}